\documentclass[aps,pre,twocolumn,showpacs,amsfonts]{revtex4}
\usepackage{epsfig}
\usepackage{graphicx}
\usepackage{dcolumn}
\usepackage{bm}

\begin{document}

\title{Dynamics of threads and polymers in turbulence: power-law distributions and synchronization}

\author{Itzhak Fouxon$^1$}
\author{Harald A. Posch$^{2}$}

\affiliation{$^1$ Raymond and Beverly Sackler School of Physics and Astronomy,
Tel-Aviv University, Tel-Aviv 69978, Israel}
\affiliation{$^2$ Computational Physics Group, Faculty of Physics, University of Vienna, Boltzmanngasse 5, A-1090 Wien, Austria}

\date{\today }

\begin{abstract}

We study the behavior of threads and polymers in a turbulent flow. These objects have finite spatial extension, so the flow along them
differs slightly. The corresponding drag forces produce a finite average stretching and the thread is stretched most of the time. Nevertheless,
the probability of shrinking fluctuations is significant and is known to decay only as a power-law. We show that
the exponent of the power law is a universal number independent of the statistics of the flow. For polymers the coil-stretch transition exists:
the flow must have a sufficiently large Lyapunov exponent to overcome the elastic resistance and stretch the polymer from the coiled state it takes otherwise. The probability of shrinking from the stretched state
above the transition again obeys a power law but with a non-universal exponent. We show that well above the transition the exponent becomes
universal and derive the corresponding expression. Furthermore, we demonstrate synchronization: the end-to-end distances of threads or polymers
above the transition are synchronized by the flow and become identical. Thus, the transition from Newtonian to non-Newtonian behavior in dilute polymer solutions can be seen as an ordering transition.

%

\end{abstract}


\maketitle

\section{Introduction}

Lagrangian chaos of the motion of fluid particles may lead to a non-trivial behavior of small objects dragged by the flow,
because the exponential separation of nearby fluid particles, that holds in chaos, stretches different parts
of small objects apart. While this is of no consequence  if the immersed object is a solid, for elastic objects the stretching is important.
Here we address the behavior of threads and of polymers immersed in a turbulent flow. Threads are defined as objects that resist the increase of their end-to-end distance only when the latter is close to the maximal length of the thread. For polymers, in contrast, the resistance exists at all distances, but it is only linear in a wide range of scales much smaller than the maximal length. The distribution of the end-to-end
distances of these objects obeys a power law \cite{BFL1,Chertkov,Th,Celani,Afonso,St,GW}, while the exponent of the power law is non-universal \cite{BFL1}. Here we derive a new region of universality for the exponent.
We also show that the flow synchronizes different particles within spatial
domains with extensions comparable to the Kolmogorov length, effectively making the vectors of the end-to-end distances
to become equal.

The main application of our results is to turbulence, where the positive Lyapunov exponent $\lambda$ is estimated as a characteristic
value of the modulus of the velocity gradient \cite{review} and Lagrangian chaos holds. The details of the statistics of the turbulent velocity
field are unknown, so one could expect that not much can be said on the behavior of the immersed threads and polymers. Nevertheless, we show that both a power-law behavior and synchronization hold. This is thanks to the universal nature of the derivation, which makes only quite
general assumptions on the statistics of the flow, that are plausible for turbulence. Our results are accessible to experiment
and may be tested.

Consider, placing a thread, with maximal length much smaller than the Kolmogorov length $\eta$, in a turbulent fluid. Typically, the thread is coiled initially and
the flow will straighten it by exponentially separating the thread's ends with time. The stretching will be arrested at about the maximal length.
After that, the thread will randomly change its orientation in the flow with its end-to-end distance fluctuating near the maximal length. From the
modern understanding of the behavior of polymers in the flow described below, it may be expected that the fluctuations are strong and that the
frequency, with which small extensions occur in the flow, decays only as a power-law in the extension. Our results show that the exponent of the
corresponding power-law in the distribution of the end-to-end distances is a universal number independent of the statistics of the
velocity (cf. \cite{SL1} and the discussion in the Conclusion). Thus we provide an analytic result for any turbulent (independent of the Reynolds number) or  chaotic flow, that obeys the proper conditions of decay of correlations.

A similar situation holds for polymers. For polymers the coil-stretch transition exists: when turbulence is not sufficiently vigorous, the elastic
forces overcome polymer's stretching by turbulence and the polymer spends most of the time in the un-stretched, coil-like state. However when the
strength of turbulent fluctuations increases, the stretching by the flow overcomes the elastic resistance and the polymer is stretched \cite{Lumley1,Lumley2,deGennes}. The quantitative criterium for the transition in the real turbulent flow was obtained in \cite{BFL1}. It was
shown, with no modeling of turbulence, that the transition occurs at $\lambda\tau=1$, where $\tau$ is the polymer relaxation time, see also
\cite{St}. At $\lambda\tau<1$, the polymer is coiled, while at $\lambda\tau>1$ it is stretched. A similar criterion was obtained independently for
the white noise (Kraichnan) model of velocity in \cite{Chertkov}. It was also shown that the probability density function of the polymer size
obeys the power-law, both for real turbulence \cite{BFL1} and for the Kraichnan model \cite{Chertkov,Th,Celani,Afonso}. For turbulence, the
exponent of the power-law is non-universal, however near the coil-stretch transition it takes a universal form \cite{BFL1}. The numerical derivation of the power-law for turbulence above the transition was made in \cite{GW}. However, the exponent could not be resolved numerically.

Here we show for real turbulence, that the exponent of the power-law admits another universal limit. Namely we derive the universal expression
for the exponent well above the transition at $\lambda\tau\gg 1$, cf. \cite{Th} and the Comment below. Our prediction is accessible to the numerical test, cf. \cite{GW}.

An important property arises, when one passes from the consideration of a single particle to many particles. We show that the Lagrangian chaos orders
the particles such that their end-to-end distances become identical. This gives a very important insight into the behavior of dilute solutions of such
particles, allowing a better understanding of  the particles' back reaction on the flow. In particular, it gives some  microscopic insight into the
possibility to describe the polymer degrees of freedom above the coil-stretch transition by a single vector field, that was derived by macroscopic means
in \cite{FL,BFL2}.
The present work indicates that the respective vector, which gives the macroscopic equations of the dilute polymer solution
a form similar to magnetohydrodynamics, is simply the end-to-end vector of a single polymer, weighted by the number of polymers in the unit volume and
the elastic properties of the polymer molecule.
Furthermore, the fact that the effective form of the particles in the flow is rod-like,
where ``rods" of different particles are aligned, is important for deriving the correct criterium when the solution can be considered dilute.
It was shown in \cite{Batchelor} that for aligned rods the condition of diluteness is much milder than for randomly oriented rods (the
effective range of influence of one rod with another is ellipsoidal for the former and spherical for the latter). Here, we postpone
the thorough discussion of implications of our results for the hydrodynamics of dilute polymer solutions to future work. We only notice that the
implications are robust.

These observations seem particularly important for turbulent dilute polymer solutions. More than $60$ years ago, it was found that
even an addition of a minute amount of polymers to a fluid can significantly change its properties such as turbulent drag (``drag reduction") \cite{GB}.
Despite high practical importance of this phenomenon and much theoretical effort, the understanding of turbulence in dilute polymer solutions is still
rather limited. It was clarified, though, that this is a threshold phenomenon: turbulence must be sufficiently strong to overcome the elasticity
of individual polymers and stretch them. As the Reynolds number and $\lambda$ grow, for $\lambda$ larger than the coil-stretch transition value $1/\tau$,
the fluid becomes non-Newtonian. What our result implies then is that this transition has similarity to a magnetic phase transition. The configurations
of the polymers are uncorrelated below the transition, while all polymers ``point" in the same direction above the transition. Moreover, even below the
transition there arise isolated ordered domains, within which the polymers are stretched and aligned. The volume fraction of domains, where polymers are stretched above a certain size, obeys a power law in that size. As $\lambda\tau$ increases, these domains become more and more frequent in space, starting to occupy most of the volume as $\lambda\tau$ crosses unity.

\section{Small threads in chaotic and turbulent flows}
\label{renewal}

We first consider the dynamics of a thread carried by a turbulent flow. We assume that the thread is much smaller than the Kolmogorov length $\eta$,
so the difference of
velocities at the thread's ends can be approximated by the first term in the Taylor expansion. We study the dynamics of the
end-to-end distance $\bm R$, assuming that, as long as the thread is not stretched to about its full length $R_{max}$, the ends are simply dragged
by the flow with the local fluid velocity $\bm u$.  If $\bm x$ is the coordinate of one of the ends, then
\begin{eqnarray}&&
\!\!\!\!\!\!\!\!\!
\dot {\bm R}=\bm u(\bm x+\bm R, t)-\bm u(\bm x, t)\approx \sigma \bm R,\ \ R\ll R_{max},
\end{eqnarray}
where $\sigma_{ij}(t)$ is the matrix of velocity derivatives $\partial_j u_i$ taken in the frame of the fluid particle $\bm x(t)$. Without the
constraint $R\ll R_{max}$, the above equation is the equation governing the separation of two Lagrangian trajectories
(trajectories of fluid particles) in the flow.
Hence $R$ evolving according to the above equation will become of the order of $R_{max}$ at some $t$.
We assume that the thread's resistance to stretching can be described by a radial force and hence the equation on $\bm R$ can be
written as
\begin{eqnarray}&&
\dot {\bm R}=\bm u(\bm x+\bm R, t)-\bm u(\bm x, t)-\nabla U(R),
\end{eqnarray}
with some "potential" $U(R)$. We assume that $\nabla U(R)$ is appreciable only for $R\sim R_{max}$, where it grows indefinitely, reflecting the
resistance of the thread to further stretching. 
For example, as a model for  threads,  one may consider the FENE model of polymers \cite{JC},
\begin{eqnarray}&&
\nabla U_{FENE}(R)=\frac{\epsilon \bm R}{1-(R/R_{max})^2},
\end{eqnarray}
where the spring constant $\epsilon$ is assumed to be small, such that the thread resistance to stretching is insignificant at $R\ll R_{max}$.
Thus, a polymer with very large relaxation time (and, hence,  well above the coil-stretch transition, see below) would be considered as a thread.

Here we wish to describe the region $R\ll R_{max}$, where the statistics becomes largely independent of the details of $U(R)$.
Decomposing $\bm R=R_{max}\exp[\rho]{\hat n}$, where ${\hat n}$ is a unit vector describing the thread's orientation, one has
\begin{eqnarray}&&
\!\!\!\!\!\!
\dot {\rho}={\hat n}\sigma{\hat n}-f'(\rho),\ \ f'(\rho)\equiv R_{max}^{-1}e^{-\rho}U'\left[R_{max}\exp[\rho]\right],\nonumber\\&& \!\!\!\!\!\!
\dot {{\hat n}}=\sigma{\hat n}-{\hat n}({\hat n}\sigma{\hat n}), \label{a2}
\end{eqnarray}
where we introduced the function $f'(\rho)$ for further convenience. For the FENE model we have
\begin{eqnarray}&&
f'_{FENE}(\rho)\equiv \frac{\epsilon}{1-\exp[2\rho]},\nonumber\\&&
f_{FENE}(\rho)=-\frac{\epsilon}{2} \ln \left(\exp[-2\rho]-1\right).
\end{eqnarray}

It is essential that the dynamics of ${\hat n}$ decouples from the one of $\rho$, while $\rho$ is driven by the "external", $\rho-$independent noise ${\hat n}\sigma {\hat n}$. Furthermore, the dynamics of ${\hat n}$ is independent of $U(R)$, and is the same even without $U(R)$,
\begin{eqnarray}&&
\dot {{\hat n'}}=\sigma{\hat n'}-{\hat n'}({\hat n'}\sigma{\hat n'}),\ \ \dot{\rho'}={\hat n'}\sigma{\hat n'}. \label{a12}
\end{eqnarray}
This equation describes the evolution of the distance between two infinitesimally-close trajectories of the fluid particles. The Lyapunov
exponent $\lambda$ is the asymptotic logarithmic growth rate of the modulus of that distance,
\begin{eqnarray}&&
\lambda=\lim_{t\to\infty}\frac{1}{t}\int_0^t {\hat n'}\sigma{\hat n'}dt',
\end{eqnarray}
where the existence of the limit follows from the law of large numbers, provided $\sigma$ has a finite correlation time $\tau_c$
(which is the case for turbulence \cite{review}).

We conclude from the above that ${\hat n}\sigma {\hat n}$ in Eq.~(\ref{a2}) has a positive mean $\lambda$. Then, at $\rho\ll -1$,
where $f'(\rho)$ is negligible, the dynamics is a random motion with a mean drift due to $\lambda$. One has ${\dot \rho}=\lambda+\xi(t)$, where
$\langle \xi \rangle=0$. We designate averages by angular brackets.
From this equation one expects an  exponential behavior of the steady-state
probability density function (PDF) $P_{ss}$ of $\rho$, like the density of a gas under gravity. We first study the problem
within the so-called Kraichnan model, where the velocity field $\bm u$ is modeled as a Gaussian
random field with zero mean and zero correlation time. Most of the results for the Kraichnan model can be inferred from \cite{Chertkov,Th,Celani,Afonso}, however there is one important new observation that we make here.
The demands of statistical isotropy and homogeneity in space imply that $\sigma$
is a stationary Gaussian matrix process with zero mean and pair-correlation function \cite{review}
\begin{eqnarray}&&
\!\!\!\!\!\!\!\!\langle\sigma_{\alpha\beta}(t)\sigma_{\gamma\delta}(0)\rangle\!=\!D\delta(t)\left[(d\!+\!1)\delta_{\alpha\gamma}\delta_{\beta\delta}\!
-\!\delta_{\alpha\delta}\delta_{\beta\gamma}\!-\!\delta_{\alpha\beta}\delta_{\gamma\delta}\right]
.\nonumber
\end{eqnarray}
Here, $d$ is the space dimension.
The above form respects the incompressibility condition and can be shown to be unique under the above assumptions on the statistics of $\bm u$. For this model, ${\hat n}\sigma{\hat n}$ is statistically equivalent \cite{BF} to the sum of $\lambda=d(d-1)D/2$ and a white noise $\xi$, so the effective dynamics is (cf. \cite{Th})
\begin{eqnarray}&&
\!\!\!\!\!\!\!\!\!\!\!\!\!
{\dot \rho}=\lambda+\xi-f'(\rho),\ \ \langle \xi(t)\xi(t')\rangle=(d-1)D\delta(t-t').\label{a11}
\end{eqnarray}
Here, $D$ is a diffusion coefficient.
The associated Fokker-Planck equation \cite{Risken} gives in the steady state
\begin{eqnarray}&&
2\left[\left(\lambda-f'(\rho)\right)P_{ss}\right]'-(d-1)D P^{''}_{ss}=0. \label{a3}
\end{eqnarray}
The steady-state solution must have a constant flux $C$,
\begin{eqnarray}&&
2\left[\lambda-f'(\rho)\right]P_{ss}-(d-1)D P'_{ss}=C.
\end{eqnarray}
At $\rho\to-\infty$, one can neglect the $f'(\rho)$ term, and the solution becomes equal to a sum of $C/\lambda$ and an exponentially decaying solution.
The condition for $P_{ss}$ to be normalizable gives $C=0$, which leads to
\begin{eqnarray}&&
P_{ss}(\rho)=\frac{1}{N}\exp\left(
\frac{2\lambda \rho-2f(\rho)}{(d-1)D}\right),\nonumber\\&&
N\equiv \int_{-\infty}^0 \exp\left(\frac{2\lambda \rho'-2f(\rho')}{(d-1)D}\right)
d\rho',\label{a7}
\end{eqnarray}
where the properties of $f(\rho)$ are such that the normalization integral $N$ is determined by $R\sim R_{max}$ or $|\rho| \sim 1$.
At $\rho\ll -1$ one may neglect $f(\rho)$, and the solution is exponential. At $R\ll R_{max}$, one finds
\begin{eqnarray}&&
P_{ss}(R)\approx \frac{1}{N R_{max}}\left(\frac{R}{R_{max}}\right)^{-1+\frac{2\lambda}{(d-1)D}},\ \ R\ll R_{max}.\nonumber
\end{eqnarray}
It is tempting to consider $\lambda/D$ as a parameter in the above equation, hoping that this will allow to describe a wider class of physical
situations \cite{Th}. Nevertheless, it will be clear below that only the actual value of $2\lambda /[(d-1)D$ equal
to $d$ brings the physically meaningful result. After this substitution,
the PDF for the end-to-end distance $R$ becomes
\begin{eqnarray}&&
P_{ss}(R)\approx \frac{1}{N R_{max}}\left(\frac{R}{R_{max}}\right)^{d-1},\ \ R\ll R_{max}.\label{a222}
\end{eqnarray}
In particular, for the FENE model one finds
\begin{eqnarray}&&
P_{ss}^{FENE}(\rho)=\frac{1}{N}
\exp\left(\rho d-\frac{\rho d \epsilon}{\lambda}\right)\nonumber\\&&\times
\left(1-\exp[2\rho]\right)^{d \epsilon/[2\lambda]}.
\end{eqnarray}
For $P_{ss}(R)$ we find
\begin{eqnarray}&&
P_{ss}^{FENE}(R)=\frac{1}{N R_{max}}\left(\frac{R}{R_{max}}\right)^{d-1-\rho d \epsilon/\lambda}
\nonumber\\&&\times
\left[1-\left(\frac{R}{R_{max}}\right)^2\right]^{d \epsilon/[2\lambda]}.
\end{eqnarray}
We observe in Eq.~(\ref{a222}) that the PDF is a positive power law (the model assumes incompressibiliy
and is meaningful only for $d\geq 2$). Thus, in the limit of
zero correlation time, the PDF of the end-to-end distance decays to small $R$ rather slowly. Furthermore, the PDF (\ref{a7})  depends only weakly
on the details
of the non-elasticity expressed by $f(\rho)$ and of the strength $D$ of the velocity gradient fluctuations via a multiplicative constant. For larger $D$, the flow stretches the
thread stronger, but the fluctuations of ${\hat n}\sigma{\hat n}$ to negative values are also stronger.

We test the above predictions numerically for two and three dimensions,  using the following discretization (cf. \cite{Kivotides}). We employ the random renewal model, where
$\sigma_{\alpha\beta}(t)$ is a piecewise continuous process, $\sigma(t)=\sigma^k$ at $k\Delta t\leq t\leq (k+1)\Delta t$, and  where $\sigma^k$ are random, independent Gaussian matrices with zero mean and pair correlation function
\begin{eqnarray}&&
\langle \sigma^k_{\alpha\beta}\sigma^{l}_{\gamma\delta}\rangle=\frac{\delta^{kl}D}{\Delta t}\left[(d+1)\delta_{\alpha\gamma}\delta_{\beta\delta}-\delta_{\alpha\delta}\delta_{\beta\gamma}-\delta_{\alpha\beta}
\delta_{\gamma\delta}\right].\nonumber
\end{eqnarray}
The above process reduces to Eq.~(\ref{a2}) in the limit $\epsilon_0\equiv D\Delta t\to 0$. For the three-dimensional case, the Gaussian matrix with the above correlation
function and zero mean can be generated using the ansatz
\begin{eqnarray}&&
\sigma^k_{\alpha\beta}=\frac{1}{\Delta  t}
\sqrt{\frac{\epsilon_0}{2}}\left[\sqrt{5}\epsilon_{\alpha\beta\gamma}V^k_{\gamma}+\sqrt{3}f^k_{\alpha}f^k_{\beta}-
\frac{\delta_{\alpha\beta}(f^k)^2}{\sqrt{3}}\right], \nonumber \label{a1}
\end{eqnarray}
where $\epsilon_{\alpha\beta\gamma}$ is the antisymmetric symbol, and
$\bm V^k$ and $\bm f^k$ are independent Gaussian random vectors with zero mean and pair correlations
$\langle V^k_{\alpha}V^l_{\beta}\rangle=\delta^{kl}\delta_{\alpha\beta}$ and
$\langle f^k_{\alpha}f^l_{\beta}\rangle=\delta^{kl}\delta_{\alpha\beta}$, respectively. In the two-dimensional case, the ansatz is
\begin{eqnarray}&&
\sigma^k_{\alpha\beta}=\frac{1}{\Delta  t}
\sqrt{\frac{\epsilon_0}{2}}\left[2\epsilon_{\alpha\beta 3}V^k+\sqrt{2}f^k_{\alpha}f^k_{\beta}-
\frac{\delta_{\alpha\beta}(f^k)^2}{\sqrt{2}}\right],\nonumber
\end{eqnarray}
where $V^k$ is a set of independent random variables with zero mean and pair correlation $\langle V^k V^l\rangle=\delta^{kl}$, and
$\bm f^k$ are independent two-dimensional Gaussian random vectors with zero mean and pair correlation
$\langle f^k_{\alpha}f^l_{\beta}\rangle=\delta^{kl}\delta_{\alpha\beta}$.
The above formulas make it clear that in the Kraichnan model the
antisymmetric part of $\sigma$, which describes the vorticity, and the symmetric part, which describes the strain, are independent.
We also note that in two dimensions the evolution matrix $\exp[\Delta t\sigma^k]$ could be written explicitly, by rewriting the
ansatz above with the help of Pauli matrices $\sigma_i$,
\begin{eqnarray}&&
\Delta t \sigma^k=\sqrt{\epsilon_0}f_1f_2\sigma_1+i \sqrt{2\epsilon_0}V^k\sigma_2+\sqrt{\epsilon_0}\left(f_1^2-f_2^2\right)\sigma_3/2.\nonumber
\end{eqnarray}
After $\bm V^k$ ($V^k$) and $\bm f^k$ are generated, the
evolution of the end-to-end vector  $\bm R^k=R^k{\hat n}^k$ of the thread follows from
\begin{eqnarray}&&
\bm N^{k+1}=\exp[\Delta t\sigma^k]{\hat n}^k,\ \ {\hat n}^{k+1}=\frac{\bm N^{k+1}}{N^{k+1}}
,\nonumber\\&& R^{k+1}=\min[R^kN^{k+1}, R_{max}], \label{a24}
\end{eqnarray}
where $R^kN^{k+1}=|\exp[\Delta t\sigma^k]\bm R^k|$. The last equation of (\ref{a24}) describes the simplest model of resistance to stretching:
the thread is stretched as if its ends are two fluid particles, unless the distance
between the ends reaches $R_{max}$, at which value the distance growth is arrested. This corresponds to a potential $U(R)$, which rises sharply
near $R_{max}$. Hence,  one may expect that the power law will extend up to values close to $R_{max}$.

\begin{figure}
\includegraphics[width=8.0 cm,clip=]{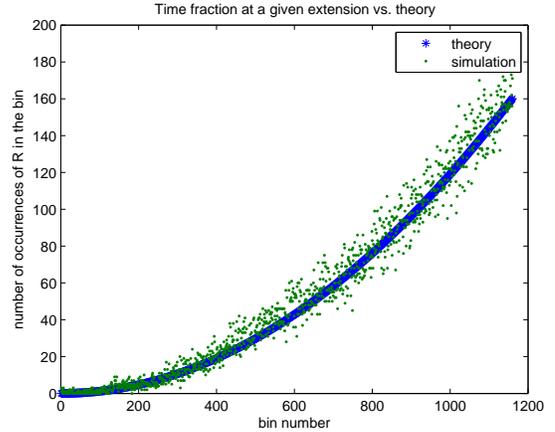}
\caption{Steady-state distribution for the end-to-end distance $R$ of a thread immersed in a turbulent fluid in three dimensions. $R$ on the abscissa
is given in multiples of the bin width, 1/1200. The distribution is not normalized. The smooth line is a parabolic fit,
as predicted by Eq. (\ref{a222}) for $d = 3$, to the experimental points.}
\label{Histogram}
\end{figure}

The numerical result of $10^5$ iterations of the above dynamics in three dimensions
(where, without loss of generality, we use reduced units, for which  $R_{max}=1$ and $\epsilon_0\equiv D\Delta t = 0.02$) is shown in the
histogram of  Figure \ref{Histogram}. There, the interval $0 < R < 1$ is split into $1200$ equally-spaced bins labeled by $i$, and the number of
occurrences of $R^k$ in each bin is counted. This sum is  proportional to the fraction of time that $R$ spends in the respective bin, for which the theory, Eq. (\ref{a222}), predicts a parabolic dependence on $R$, for $R$ sufficiently smaller than $R_{max}=1$. Removing $50$ bins near $R=R_{max}=1$, the resulting histogram shown in Figure \ref{Histogram} fits the parabola $C i^2, i \in \{1,2,\dots,1150\} $ for $C=1.2\times 10^{-4}$ very well
(the scattered points deviate from the fit by less than $10$ \%). The power law even extends to values of $R$ very close to $R_{max}$ in
accord with the expectation mentioned above.
Analogous results for simulations in two dimensions are shown in the histogram of Figure \ref{TwoDimensionalSimulations}, where $2 \times 10^5$ iterations were made. A very satisfactory fit to a linear profile, as predicted by  Eq. ({\ref{a222}) for $d = 2$, is shown by the straight line, where the scatter of the
experimental points does not exceed  $10$ \%.
\begin{figure}
\includegraphics[width=8.0 cm,clip=]{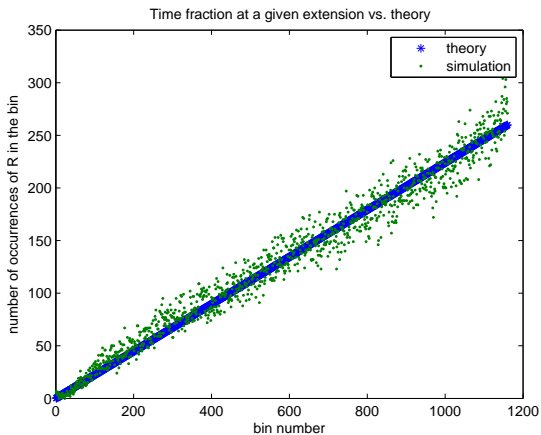}
\caption{Steady-state distribution for the end-to-end distance $R$ of a thread immersed in a turbulent fluid in two dimensions. $R$ on the abscissa
is given in multiples of the bin width, 1/1200. The distribution is not normalized. The smooth line is a linear fit,  as
predicted by Eq. (\ref{a222}) for $d = 2$, to the experimental points.}
 \label{TwoDimensionalSimulations}
\end{figure}
The physical significance of the power law is illustrated in  Figure \ref{ThreadSizeFluctuations} for the three-dimensional case.
It shows a realization of $R^k$ in the steady state over $1000$ steps of its evolution. It is clear that the excursions to small $R$ occur quite often, as the slow power-law decay of $P_{ss}(R)$ to small $R$ would make one to expect.
\begin{figure}
\includegraphics[width=8.0 cm,clip=]{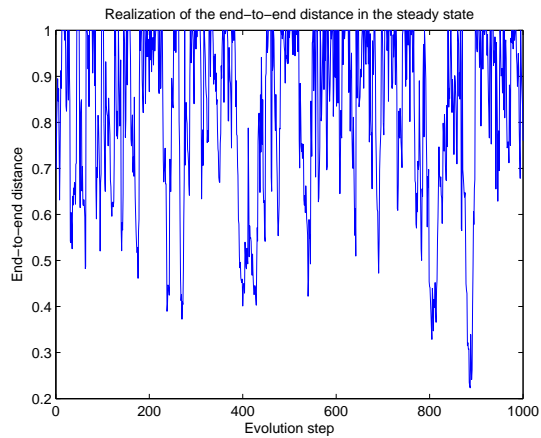}
\caption{Shown is a time series  $R^k$ for 1000 time steps of a thread in three dimensions. The excursions to $R$ significantly smaller than $R_{max}=1$ are quite frequent.}
\label{ThreadSizeFluctuations}
\end{figure}

It is possible to generalize the above results to any finite-correlated $\sigma(t)$, in particular, to the
one of turbulence. Such a generalization is very important, as it gives an experimentally testable prediction for the behavior of threads in a
chaotic flow and, in particular, in turbulence. It requires to take the Lyapunov instability of the flow into account, which involves  the time
evolution of infinitesimally-close trajectories. This is considered in the following section.

\section{Separation vector between two infinitesimally-close trajectories in a chaotic flow: a new result}

Below we derive two results on the behavior of the separation vector between two infinitesimally-close trajectories in a chaotic flow. The first result is new, while the second one is less known. These results
are at the basis of several physical predictions obtained in the following sections. Where there is no ambiguity, we
use the same notation as used in other sections for other quantities.

\subsection{Relaxation of the orientation of different separation vectors to the same vector}
\label{General}

It follows from the previous sections that the distance $\bm R$ between two infinitesimally close trajectories obeys
\begin{eqnarray}&&
\frac{d\bm R}{dt}=\sigma\bm R,\ \ \bm R(t)=W(t)\bm R(0),  \label{a12}
\end{eqnarray}
where we introduced the Jacobi matrix $W(t)$ of derivatives of fluid particles' coordinates with respect to their initial position. If $\bm q(t, \bm x)$
is a Lagrangian trajectory,
\begin{eqnarray}&&
\frac{\partial \bm q(t, \bm x)}{\partial t}=\bm u\left[t, \bm q(t, \bm x)\right],\ \ \bm q(0, \bm x)=\bm x,
\end{eqnarray}
then the Jacobi matrix describes the distance $\bm R(t)$ between two infinitesimally close trajectories starting at $\bm x$ and $\bm x+\bm R(0)$
 according to
\begin{eqnarray}&&
\bm R(t)=\bm q(t, \bm x+\bm R)-\bm q(t, \bm x)\approx W(t, \bm x)\bm R,\nonumber\\&& W_{ij}(t, \bm x)\equiv\frac{\partial q_i(t, \bm x)}
{\partial x_j}.\label{a122}
\end{eqnarray}
Below we suppress the spatial index in $W(t, \bm x)$, where it is clear from the context.
The matrix $W(t)$ obeys
\begin{eqnarray}&&
\frac{dW}{dt}=\sigma W,\ \ W(0)=1, \label{a14}
\end{eqnarray}
in accord with Eq.~(\ref{a12}). The variable $\rho(t)$ and the unit vector ${\hat n}(t)$ defined by $\bm R(t)=R(0)\exp[\rho]{\hat n}(t)$ satisfy
\begin{eqnarray}&&
\!\!\!\!\!\!\!\!\!\frac{d{\hat n}}{dt}=\sigma {\hat n}-{\hat n}\left[{\hat n}\sigma {\hat n}\right],\ \
{\hat n}(t)=\frac{W(t){\hat n}(0)}{|W(t){\hat n}(0)|},\label{a2000}\\&&\!\!\!\!\!\!\!\!\!
\rho(t)=\int_0^t {\hat n}\sigma {\hat n}dt',\ \ \lim_{t\to\infty}\frac{1}{t}\int_0^t {\hat n}\sigma {\hat n}dt'=\lambda. \nonumber
\end{eqnarray}
We introduce the matrix decomposition $W(t)=R\Lambda N$, where $R$ and $N$ are orthogonal matrices, while $\Lambda$ is diaginal,
$\Lambda_{ij}=\delta_{ij}\exp[\lambda_i(t) t]$.
According to the Oseledec theorem \cite{Oseledets},  $\lim_{t\to\infty}\lambda_i(t)=\lambda_i$, where $\lambda_i$ are the Lyapunov exponents,
while $N(t)$, diagonalizing $W^TW$, saturates at large times at a constant matrix $N^{\infty}$. Here, the index $T$ means transposition.
We assume that the Lyapunov exponents are ordered, $\lambda_i(t)\geq \lambda_{i+1}(t)$, and that $\lambda_1$ is strictly greater than $\lambda_2$.
We have
\begin{eqnarray}&&
[W(t){\hat n}(0)]_i=R_{ij}\exp[\lambda_j(t) t]N_{jk}{\hat n}_k(0),
\end{eqnarray}
where the summation over repeated indices is presumed. At times, such that $\exp[\lambda_1(t)t]\gg \exp[\lambda_2(t)t]$, the $j=1$ term dominates
the sum over $j$. Introducing the unit vectors ${\hat m}_i\equiv R_{i1}$ and ${\hat l}_i\equiv N_{1i}$, we find that
\begin{eqnarray}&&
W(t){\hat n}(0)\approx \exp[\lambda_1(t) t]{\hat m}(t)\left[{\hat l}\cdot {\hat n}(0)\right],
\nonumber\\&&
{\hat n}(t)=\frac{W(t){\hat n}(0)}{|W(t){\hat n}(0)|}\approx {\hat m}(t) sign \left[{\hat l}\cdot {\hat n}(0)\right]
, \label{a44}
\end{eqnarray}
where the sign factor corresponds to interchanging the labels of the trajectories and has no physical significance for  the applications we consider in
the following sections. The approximation above holds for all ${\hat n}(0)$, for which ${\hat l}\cdot {\hat n}(0)$ is much larger than the
small parameter $\exp([\lambda_1(t)-\lambda_2(t)]t)$. The Oseledets theorem implies that ${\hat l}$ saturates at a constant vector ${\hat l}_0$
at large times \cite{Oseledets}. This can be verified directly by deriving from the equation for $W$ the following equation for ${\hat l}$, see \cite{FL},
\begin{eqnarray}&&
\frac{d{\hat l}_i}{dt}=\sum_{j=2}^d\frac{\left(R^T\sigma R\right)_{1j}+\left(R^T\sigma R\right)_{j1}}{\sinh\left[t\left(\lambda_1(t)-\lambda_j(t)
\right)\right]}N_{j1}.
\end{eqnarray}
At times considered in Eq.~(\ref{a44}), the time derivative of ${\hat l}$ decays exponentially, which indicates that ${\hat l}$ is approximately constant
at those times. Thus the approximation (\ref{a44}) holds for all initial conditions besides the
$\exp[(\lambda_2-\lambda_1)t]$ vicinity of the plane ${\hat l}_0\cdot {\hat n}(0)=0$.

Choosing with no loss the sign of ${\hat n}(0)$ so that ${\hat l}_0\cdot {\hat n}(0)>0$, we find from Eq.~(\ref{a44}) that
\begin{eqnarray}&&
\!\!\!\!\!\!\!{\hat n}(t)\approx {\hat m}(t)
,\ \ \exp[(\lambda_2-\lambda_1)t]\ll \min[1, {\hat l}_0\cdot {\hat n}(0)]
. \label{a14}
\end{eqnarray}
It is possible to verify from the equation for $R(t)$ that ${\hat m}(t)$ obeys the same equation as ${\hat n}(t)$, so the above approximation is
consistent \cite{FL}. Since the corrections to Eq.~(\ref{a14}) are of order $\exp[t\left(\lambda_1-\lambda_2\right)]$, we conclude that different
${\hat n}(t)$, obtained from different initial conditions, exponentially relax to the same unit vector ${\hat m}$.

The most remarkable feature of Eq.~(\ref{a14}) is that ${\hat n}(t)$ lost all dependence on the initial condition ${\hat n}(0)$,
provided ${\hat n}(0)$ is not in the narrow vicinity of the plane ${\hat l}_0\cdot {\hat n}(0)=0$. Thus, any initial condition outside the latter
plane will relax to ${\hat m}(t)$ at sufficiently large times.

The observation of this property seems new, and we consider it now in more detail.
We introduce a unit vector ${\hat q}(t)$ by ${\hat q}(t)\equiv N(t){\hat n}(0)$. The Oseledets theorem ensures that ${\hat q}(t)$ saturates
at a constant vector $\bm q^{\infty}\equiv N^{\infty}{\hat n}(0)$ in the limit $t\to\infty$. The following exact relation follows from
Eq.~(\ref{a2000}),
\begin{eqnarray}&&
R^T(t){\hat n}(t)=\frac{\Lambda(t) {\hat q}(t)}{|\Lambda {\hat q}(t)|}=
\Biggl(\frac{q_1}{Q}, \frac{q_2}{Q}\exp[t\left(\lambda_2(t)-\lambda_1(t)\right)],
\nonumber\\&&
\ldots, \frac{q_d}{Q}\exp[t\left(\lambda_d(t)-\lambda_1(t)\right)]\Biggr),
\end{eqnarray}
where we defined
\begin{eqnarray}&&
Q^2\equiv q_1^2+q_2^2\exp[2t\left(\lambda_2(t)-\lambda_1(t)\right)] +\ldots \nonumber
\\&& +\exp[2t\left(\lambda_d(t)-\lambda_1(t)\right)]q_d^2.\nonumber
\end{eqnarray}
We find that for such ${\hat n}(0)$, for which $q^{\infty}_1\neq 0$, we have
\begin{eqnarray}&&
\lim_{t\to\infty} R^T(t){\hat n}(t)=(1, 0, \ldots, 0),
\end{eqnarray}
where it is assumed that the initial choice of the labels of the trajectories corresponds to the positive sign of ${\hat l}_0\cdot {\hat n}(0)$
(otherwise there appears $-1$ in the vector on the RHS above instead of $1$). Since $m(t)=R(1, 0, \ldots, 0)$, the above equation gives the more
precise formulation of Eq.~(\ref{a14}).
The condition $q^{\infty}_1\neq 0$ means that the limit holds for all ${\hat n}(t)$ such that ${\hat n}(0)$ obeys
${\hat l}_0\cdot {\hat n}(0)\neq 0$. This limit  holds for almost every trajectory, with the possible exception of a set of trajectories with a vanishing volume
of their initial conditions.

\subsection{A relation for the exponential growth of the distance between two infinitesimally-close trajectories}

It follows from the Oseledets theorem that the separation $R(t)$ between two infinitesimally close trajectories behaves exponentially,
\begin{eqnarray}&&
\lim_{t\to\infty}\frac{1}{t}\ln\left(\frac{R(t)}{R(0)}\right)=\lambda. \label{a449}
\end{eqnarray}
For the case of incompressible flow considered here, the above limit holds for any initial position of trajectories, possibly except for a set of
positions with zero spatial volume. Thus, one could expect $R^{\alpha}(t)$ to behave as $\exp[\alpha\lambda t]$. This expectation fails due to the very
intermittent nature of the growth of the moments. There are small regions in space, for which $R(t)$ grows with a time-dependent
(local) exponent larger than $\lambda$,  and other
regions, where $R(t)$ grows with a time-dependent local  exponent smaller than $\lambda$, or even decays exponentially. The volume of these regions exponentially decays to zero as $t\to\infty$, and it can be described as follows. One introduces $x(t)=\ln\left(\frac{R(t)}{R(0)}\right)/t$ and describes it by the PDF
$P(x, t)$, where the probability is defined by the fraction of the volume for which $x(t)=x$. Then the exponential decay of the volumes, for which $x(t)$
differs from $\lambda$, is described by
\begin{eqnarray}&&
P(x, t)\sim \exp\left[-tS(x-\lambda)\right],
\end{eqnarray}where $S(x)$ is the rate function, which is convex and positive everywhere except at zero, see Ref.  \cite{review} and the references therein.
Although for  large times  the volume of the regions with $x\neq \lambda$ is exponentially small, these regions still contribute significantly to the spatial
averages of $R^{\alpha}(t)$. This is, because for these regions $R^{\alpha}(t)$ can be exponentially larger than $\exp[\alpha\lambda t]$ (or smaller as
may be needed for $\alpha<0$). The resulting non-trivial statistics is described by the exponential growth
function $\gamma(\alpha)$ defined by \cite{Zeldovich,review},
\begin{eqnarray}&&
\!\!\!\!\!\!\!\gamma(\alpha)\equiv \lim_{t\to\infty} \frac{1}{t}\ln \langle R^{\alpha}(t)\rangle= \lim_{t\to\infty} \frac{1}{t}\ln \langle
|W(t){\hat n}|^{\alpha}\rangle, \label{defgam}
\end{eqnarray}
where ${\hat n}={\hat n}(0)$.
Using the cumulant expansion, one can show that $\gamma(\alpha)$ is well-defined, and using $P(x, t)$, one can see
that $\gamma(\alpha)$ is the Legendre transform of the rate function $S(x)$.
The Lyapunov exponent enters $\gamma(\alpha)$ via the relation $\gamma'(0)=\lambda$, as can be easily inferred from the definition.
Furthermore, it follows from the H\"older inequality that $\gamma(\alpha)$ is convex (which is equivalent to the convexity of $S(x)$).

Restoring the spatial argument of $W(t)$, see Eq.~(\ref{a122}), one can write down the average explicitly,
\begin{eqnarray}&&
\langle |W(t){\hat n}|^{\alpha}\rangle\equiv \int \frac{d\bm x}{\Omega}|({\hat n}\cdot\nabla_{\bm x})\bm q(t, \bm x)|^{\alpha},
\end{eqnarray}
where $\Omega$ is the system volume. For isotropic statistics the average above is independent of ${\hat n}$. Since the statistics
of $\sigma$ is determined by the small-scale turbulence, where the statistics gets isotropic, we assume below that isotropy
holds.

We consider the integral of $|W{\hat n}|^{-d-\mu}$ over the directions ${\hat n}$ and make the following change of variables in the
expression for the average,
${\hat n}'=W{\hat n}/|W{\hat n}|$. This transformation was introduced in Ref. \cite{Zeldovich}, where it was shown that the Jacobian of this
transformation of the unit sphere is $|W{\hat n}|^{-d}$, i. e. $d{\hat n}'=|W{\hat n}|^{-d}d{\hat n}$. We find
\begin{eqnarray}&&
\int \frac{d{\hat n}}{|W(t){\hat n}|^{d+\mu}}=
\int |W^{-1}(t){\hat n}'|^{\mu}d{\hat n}',
\end{eqnarray}
where we note that $|W^{-1}{\hat n}'|=1/|W{\hat n}|$. Averaging that  relation,  which holds at every point [remember that
$W_{ij}(t, \bm x)=\partial_j q_i(t, \bm x)$], and using the fact that  the average is independent of the direction, we obtain
\begin{eqnarray}&&
\langle |W(t){\hat n}|^{-d-\mu}\rangle=\langle |W^{-1}(t){\hat n}'|^{\mu}\rangle.
\end{eqnarray}
The above relation allows to draw an important conclusion for $\gamma(\alpha)$. First, putting $\mu=0$, we recover the relation
$\langle |W(t){\hat n}|^{-d}\rangle=1$, that was established in \cite{Zeldovich}. Second, differentiating the above relation
with respect to $\mu$ and substituting $\mu=0$ in the resulting relation, we obtain
\begin{eqnarray}&&
\langle |W(t){\hat n}|^{-d}\ln |W(t){\hat n}|\rangle=-\langle \ln |W^{-1}(t){\hat n}|\rangle.
\end{eqnarray}
Using the definition (\ref{defgam}) of $\gamma(\alpha)$ to write down $\gamma'(-d)$, and then using the relations above, we obtain
\begin{eqnarray}&&
\gamma'(-d)=\lim_{t\to\infty}\frac{1}{t}\frac{\langle |W(t){\hat n}|^{-d}\ln |W(t){\hat n}|\rangle}{\langle |W(t){\hat n}|^{-d}\rangle}
\nonumber\\&&
=-\lim_{t\to\infty}\frac{1}{t}\langle \ln |W^{-1}(t){\hat n}|\rangle.
\end{eqnarray}
However, the last term is nothing but $\lambda_d$, as the average of the logarithm is determined by $\lambda_d(t)$ due to the appearance of $W^{-1}$
in the average. We find
\begin{eqnarray}&&
\gamma'(-d)=\lambda_d.
\end{eqnarray}
This  relation can be found in the Appendix of \cite{FL}. It is less known but important for our analysis below.

We now summarize the properties of $\gamma(\alpha)$, that will be important in the following. This function has
two zeros, $\gamma(0)=\gamma(-d)=0$. The derivatives of $\gamma(\alpha)$ at these zeros are given by $\gamma'(0)=\lambda_1$ and
$\gamma'(-d)=\lambda_d$. These are positive and negative, respectively, provided the system
is non-degenerate and $\lambda_1>0$ (then incompressibility implies that $\lambda_d<0$ due to $\sum \lambda_i=0$). It follows then from the convexity
that $\gamma(\alpha)<0$ for $-d<\alpha<0$, and   $\gamma(\alpha)>0$ for $\alpha>0$ and $\alpha<-d$.

\subsection{The exponential growth rate function for the Kraichnan model}

The expression for $\gamma(\alpha)$ within the Kraichnan model is well-known, see \cite{review} and references therein.
Here, to keep the exposition self-contained,  we rederive the result. Using the representation $\bm R(t)=R(0)\exp[\rho(t)]{\hat n}(t)$,
where $|{\hat n}|=1$,  we find that the definition (\ref{defgam}) of $\gamma(\alpha)$ admits the following form:
\begin{eqnarray}&&
\!\!\!\!\!\!\!\!\!\!\!\!\gamma(\alpha)\equiv \lim_{t\to\infty} \frac{1}{t}\ln \left\langle \exp\left[\alpha\int_0^t {\hat n}(t')\sigma(t'){\hat n}(t') dt'\right]
\right\rangle.
\end{eqnarray}
For the Kraichnan model ${\hat n}(t')\sigma(t'){\hat n}(t')$ is Gaussian, and it follows that $\gamma(\alpha)$ is parabolic.
From $\gamma(-d)=\gamma(0)=0$ and $\gamma'(0)=\lambda_1=\lambda$ one infers that
\begin{eqnarray}&&
\gamma(\alpha)=\frac{\lambda\alpha(\alpha+d)}{d}. \label{gammaKraich}
\end{eqnarray}
Note that for the Kraichnan model $\lambda_d=\gamma'(-d)=-\lambda_1$.

\section{Derivation of $P_{ss}(R)\propto R^{d-1}$ for real turbulence}

We now generalize the result on the power-law distribution of $R$ at $R\ll R_{max}$ to any finite-correlated $\sigma(t)$, in particular, to the
one of turbulence. This gives an experimentally testable prediction on the distribution of the end-to-end distance of a small thread in turbulence.
We then use the technique introduced in \cite{BFL1} for a study of the distribution of polymers in turbulence.

At $\rho\ll -1$, over a not too large time interval $t$ such that $\rho(\tau)\ll -1$ for $0\leq \tau\leq t$, we have
\begin{eqnarray}&&
\rho(t)=\rho(0)+z(t),\ \ z(t)\equiv \int_0^t {\hat n}\sigma{\hat n} dt'. \label{a8}
\end{eqnarray}
Generally speaking, $z(t)$ grows with time, and at a sufficiently large $t$ one can neglect the contribution to $z(t)$ of a $\tau_c$ neighborhood of $t=0$.
Then $z(t)$ is approximately independent of $\rho(0)$ that, by causality, is determined only by $\sigma(t)$ with $t<0$. It follows that the
PDF of $\rho(0)+z(t)$ is given by the convolution of PDFs of $\rho(0)$ and $z(t)$. In the steady state both the distribution of $\rho(t)$ and of
$\rho(0)$ are given by $P_{ss}(\rho)$. It follows that $P_{ss}(\rho)$ must obey the following stationarity condition,
\begin{eqnarray}&&
P_{ss}(\rho)=\langle P_{ss}\left[\rho-z(t)\right]\rangle.
\end{eqnarray}
Using Laplace transforms,  one can show that the above equation has exponential solutions $\propto\exp[\mu\rho]$, where $\mu$ obeys (cf. \cite{BFL1})
\begin{eqnarray}&&
\left\langle \exp\left[-\mu \int_0^t {\hat n}\sigma{\hat n} dt'\right]\right\rangle=1.
\end{eqnarray}
It follows from the analysis in the previous section that $\mu$ obeys $\gamma(-\mu)=0$. There are two solutions, which are the same as in
the Kraichnan model: $\mu=0$ corresponding to $P_{ss}(\rho)=const$ and $\mu=d$ corresponding to $P_{ss}\propto \exp[\rho d]$. Only the latter
solution is normalizable at $\rho\to -\infty$. Passing to the variable $R$, we obtain
\begin{eqnarray}&&\!\!\!\!\!\!\!\!\!\!
P_{ss}(R)\approx \frac{1}{{\tilde N} R_{max}}\left(\frac{R}{R_{max}}\right)^{d-1},\ \ R\ll R_{max},\label{a2222}
\end{eqnarray}
where ${\tilde N}$ is a constant determined by matching the solution above to its form at $R\sim R_{max}$. The latter cannot be obtained for
a general $\sigma$. We stress that although Eq.~(\ref{a2222}) gives the same result as in the Kraichnan model, this does not signify that an effective
white noise description is possible in the region $R\ll R_{max}$, as it will become more clear in the next sections. Furthermore, the Kraichnan
model gives the correct result only for the ratio $\lambda/D$ that holds for the full model. This is most essential for the understanding of the
applicability of the model, cf. \cite{Th}.

Thus, the dynamics of a small thread thrown into a turbulent flow consists of  stretching to its full length and
rotation as a rigid rod, which is intermittently interrupted by the  shrinking and relaxation of the thread. The shrinking intervals - excursions to the domain
$R<R_{max}$ - occur rather often in accord with the power-law distribution proportional to $R^{d-1}$. To check the theoretical prediction, one needs
to measure the PDF of a small thread in a three-dimensional turbulent (or  chaotic)  flow and compare it with an  $R^2$-distribution, or make the measurement
in two-dimensions,  and compare the result with $R$.

\section{Synchronization of threads by the flow}
\label{synchro}

Having described the distribution of the end-to-end distance of a single thread, let us compare the behavior of different threads in the same velocity
gradient $\sigma$. For turbulence this means that the threads are separated by a distance much smaller than the Kolmogorov length $\eta$, but not too small,
so that the hydrodynamic interactions between the threads are still negligible. We consider the dynamics of the orientation vector  ${\hat n}$.
As the dynamics of ${\hat n}$ is the same as for two fluid particles, see Eq.~(\ref{a12}), we find immediately from Eq.~(\ref{a14})
that orientations of different threads at large times are equal to the same vector,
\begin{eqnarray}&&
\!\!\!\!\!\!\!{\hat n}(t)\approx {\hat m}(t)
. \label{a144}
\end{eqnarray}
Here, we omit the sign factor [see Eq.~(\ref{a14})], as a change in the sign of ${\hat n}$ does not change the configuration of the physical thread
in space. Thus, threads with initially different orientations relax to the same vector exponentially. As it is clear from our considerations in Section
\ref{General}, the exponent of the relaxation can be estimated as $\lambda_1-\lambda_2$.


One can expect that not only the orientations,  but also the extensions of the threads relax to the same value, such that  the flow synchronizes the
threads completely, and the end-to-end
vectors become equal for different threads in the same flow. Consider the radial variables $\rho_1$ and $\rho_2$, each of which obeys the first of
Eqs.~(\ref{a2}). Due to the equalization of orientations, at sufficiently large times the dynamics reduces to
\begin{eqnarray}&&
\dot {\rho}_i={\hat m}\sigma{\hat m}-f'(\rho_i),
\end{eqnarray}
so that - if we introduce $\delta \rho=\rho_2-\rho_1$ - we obtain a system of two equations,
\begin{eqnarray}&&
\frac{d\delta \rho}{dt}=f'(\rho_1)-f'(\rho_1+\delta\rho),\ \ \dot {\rho}_1={\hat m}\sigma{\hat m}-f'(\rho_1). \nonumber
\end{eqnarray}
This system may be considered as a Langevin-like system of equations, with an obvious-steady state solution
$P_{ss}(\delta\rho, \rho_1)=\delta\left(\delta\rho\right)P_{ss}(\rho_1)$, where the steady-state solution for a single radial variable,
$P_{ss}(\rho)$, was discussed before. One expects that this solution provides the unique steady-state solution for the system, signifying complete
synchronization between different threads. For example, for the FENE model one has
\begin{eqnarray}&&
\frac{d\delta \rho}{dt}=\epsilon\left[\frac{1}{1-\exp[2\rho_1]}-\frac{1}{1-\exp[2\rho_1+2\delta \rho]}\right].\nonumber
\end{eqnarray}
The time derivative on the RHS always has the sign opposite to that of $\delta \rho$ (remember that both $\rho_1$ and $\rho_2=\rho_1+\delta\rho$ are
negative). The above equation describes monotonous relaxation of
$\delta\rho$ to $0$, which demonstrates synchronization for the FENE model. We also verified the synchronization numerically for the
three-dimensional renewal model of Sec. \ref{renewal}. We constructed the Jacobi matrix for $n$ evolution steps,
\begin{eqnarray}&&
\!\!\!\!\!\!\!\!\!
W_n\!\!=\!\!\prod_{k=1}^n\exp\Biggl[\sqrt{\frac{5\epsilon}{2}}\epsilon_{\alpha\beta\gamma}V^k_{\gamma}+
\sqrt{\frac{3\epsilon}{2}}f^k_{\alpha}f^k_{\beta}-\sqrt{\frac{\epsilon}{6}}\delta_{\alpha\beta}(f^k)^2\Biggr],
\nonumber
\end{eqnarray}
and verified that
\begin{eqnarray}&&
\frac{[W_n W_n^T]_{ij}}{{\hat n}_i{\hat n}_j\exp[\rho_1]}\approx \frac{R_{i1}R_{j1}}{{\hat n}_i{\hat n}_j},
\end{eqnarray}
relaxes to a matrix composed only of ones for different initial conditions of ${\hat n}$. The relaxation of all elements to unity (up to order
$10^{-4}$) was already observed after $50-100$ evolution steps,  slightly depending on the initial conditions. This verifies the
alignment of different threads as described by Eq.~(\ref{a144}).
Then we verified that, with the same number of evolution steps, the values of the radial
component $\rho$ relax to the same value for different initial conditions.
Since for the Kraichnan model for $d=3$ one has $\lambda_2=0$ and $\lambda_1-\lambda_2=\lambda=3D$, see Ref. \cite{review}, which in our
dimensionless units equals $3\epsilon$ with $\epsilon=0.02$, the convergence occurs within the expected time-scale $(\lambda_1-\lambda_2)^{-1}$.
It seems highly unlikely that our conclusion on the relaxation of the radial coordinate is model-dependent, and we will assume that
it holds for any physically meaningful $f(\rho)$. Thus, our analysis brings the conclusion that different threads get synchronized by the flow.
This result can be tested experimentally.

\section{The extension distribution for a single polymer above the coil-stretch transition}

As mentioned above, threads can be considered as the limit of a polymer with a very large relaxation time, which is the case of the FENE model for the
thread. One may expect that the results are generalizable to polymers with an arbitrary relaxation time, which we pass to show.
The main difference of polymers from threads is that the former exert a significant elastic resistance force, which opposes stretching
already for $R\ll R_{max}$. In particular,
this force may arrest the stretching by the flow, which corresponds to the regime below the coil-stretch transition \cite{Lumley1,Lumley2,deGennes,BFL1}. The contents of this section and the next one are largely inferrable from the previous work and are introduced
to keep the exposition self-contained.

The results for threads can be carried over to polymers above the coil-stretch transition rather straightforwardly.
The following effective equation holds for the end-to-end vector  $\bm R$ of the polymer (see e. g. Ref.  \cite{BFL1,Chertkov}),
\begin{eqnarray}&&
\dot{\bm R}=\sigma \bm R-\frac{\bm R}{\tau}-\nabla U(R)+\bm \zeta,\label{a6}
\end{eqnarray}
where $\tau$ is the polymer relaxation time, and $\nabla U(R)$ is negligible for a wide range of scales $R_{coil}\ll R\ll R_{max}$. The white noise $\bm \zeta$
produces equilibrium fluctuations, that govern the behavior of the polymer in the coiled state  where $R\sim R_{coil}\ll R_{max}$.
This noise can be neglected  for $\lambda\tau>1$ above the coil-stretch transition
(see Ref. \cite{BFL1}), where the polymer is stretched to $R\gg R_{coil}$ and $\bm \zeta$ is small  in comparison with the term $\bm R/\tau$.
The self-consistency of this omission is checked by dropping $\bm \zeta$ in the equation above and passing again to the same variables $\rho$ and
${\hat n}$ introduced before for threads (with no ambiguity we use here the same notation).
One finds
\begin{eqnarray}&&
\!\!\!\!\!\!
\dot {\rho}={\hat n}\sigma{\hat n}-\frac{1}{\tau}-f'(\rho),\ \ \dot {{\hat n}}=\sigma{\hat n}-{\hat n}({\hat n}\sigma{\hat n}). \label{a22}
\end{eqnarray}
Like for threads, the dynamics of the orientations is the same as for a vector separating two fluid particles.
Thus, the only change in comparison with threads is that the Lyapunov exponent should be substituted by the effective value $\lambda-\tau^{-1}$.
For $\lambda\tau<1$, this change is essential, as the above equation - on average - describes a $\rho$ that decreases for $R\gg R_{coil}$, giving inconsistency
if $\bm \zeta$ is neglected. Above the coil-stretch transition, however, at $\lambda\tau>1$, the neglect is justified. In the Kraichnan model \cite{Chertkov,Th,Celani,Afonso}, the
straightforward generalization of the analysis for threads yields for $R\ll R_{max}$
\begin{eqnarray}&&
\!\!\!\!\!\!\!\!\!P_{ss}(R)=\frac{1}{ N' R_{max}}\left(\frac{R}{R_{max}}\right)^{\alpha-1}\!\!\!\!\!, \ \alpha\equiv d\left[1-\frac{1}{\lambda\tau}\right],
\label{a5}
\end{eqnarray}
where $N'$ is a constant, which should be determined by matching the above solution to the asymptotic region $R\sim R_{max}$.
The above result is the same as for the FENE model of the thread with $\epsilon=1/\tau$. This is necessary, because the calculation for the threads did not
use the smallness of $\epsilon$. Below the coil-stretch transition,  $\alpha<0$ and the above PDF is not normalizable at small $R$, which indicates the
inconsistency of neglecting $\bm \zeta$. The power law below the transition was derived in Refs. \cite{BFL1,Chertkov}.
In contrast, above the coil-stretch transition, $\alpha>0$ and the distribution is normalizable at small $R$, indicating that the normalization is determined by
$R\sim R_{max}$, where the polymer spends most of the time. Since the derivation of the power law is based on considerations local in $R$, it is readily
concluded that the power law holds at $R_{coil}\ll R\ll R_{max}$.

A noticeable difference from the case of the thread is that for polymers the exponent of the power law is non-universal and depends on the parameter $D$
of the Kraichnan model. This is due to the presence of the dimensionless parameter $\lambda\tau$. It should also be noticed that even above the
coil-stretch transition,  $P_{ss}(R)$ can be a decreasing function of $R$, which has a normalizable singularity at $R=0$.

\section{The generalization to the case of real turbulence}

The generalization to the arbitrary statistics of $\sigma$, in particular  to the one of turbulence, is a straightforward repetition of the analysis made
previously for the threads and it follows \cite{BFL1}, see also \cite{Bof}. One has at $\rho\ll -1$ that
\begin{eqnarray}&&
\rho(t)=\rho(0)+z(t)-\frac{t}{\tau},\ \ z(t)\equiv \int_0^t {\hat n}\sigma{\hat n} dt'. \label{a8}
\end{eqnarray}
Again, for sufficiently large $t$, one neglects the contribution to $z(t)$ of a $\tau_c$ neighborhood of $t=0$ and finds the following condition for
the steady-state distribution $P_{ss}(\rho)$:
\begin{eqnarray}&&
P_{ss}(\rho)=\left\langle P_{ss}\left[\rho-z(t)+\frac{t}{\tau}\right]\right\rangle.
\end{eqnarray}
Again, this equation has exponential solutions $\propto\exp[\mu\rho]$, where $\mu$ obeys (cf. \cite{BFL1})
\begin{eqnarray}&&
\left\langle \exp\left[-\mu \int_0^t {\hat n}\sigma{\hat n} dt'\right]\right\rangle= \exp\left[-\frac{\mu t}{\tau}\right].
\end{eqnarray}
The exponent $\mu$ obeys in this case the condition
\begin{eqnarray}&&
\gamma(-\mu)=-\mu/\tau. \label{stcond}
\end{eqnarray}
As before, the convexity of $\gamma(\alpha)$ (and thus of
$\gamma(-\mu)+\mu/\tau$) implies that there are just two solutions to this condition. Also again, the solution $\mu=0$ should be discarded,
as it is non-normalizable. The exponent $\mu$, however, is not universal such as in the case of the threads, because the behavior of $\gamma(\alpha)$
in general is not universal, but depends on the details of the statistics. The only general implication possible is that $\mu$ obeys $\mu<d$,
which follows from the fact that $\gamma(\alpha)$ is negative for $-d<\alpha<0$, and positive otherwise, cf. Ref.  \cite{Zeldovich}.

Thus, the steady state of the polymer molecule in the flow above the coil-stretch transition satisfies
\begin{eqnarray}&&
\!\!\!\!\!\!\!\!\!P_{ss}(R)=\frac{1}{ N'' R_{max}}\left(\frac{R}{R_{max}}\right)^{d[1-\delta]-1},
\end{eqnarray}
where $\delta=1-\mu/d>0$. For the Kraichnan model $\delta=1/(\lambda\tau)$. The condition of self-consistency of the above equation, guaranteeing
that it describes polymers above the coil-stretch transition, is $\delta<1$ which, for the Kraichnan model, reproduces the universal criterium
for the coil-stretch transition, $\lambda_1\tau=1$ (see Ref. \cite{BFL1} and below).

\section{From polymers to threads: the case of a very loose polymer}

One case, for which the general analysis provides  the exponent governing the steady state distribution $P_{ss}(\rho)$, is the thread limit
$\lambda\tau\gg 1$. We have in this case that $\mu$ is close to the solution for the thread
\begin{eqnarray}&&
\mu=d[1-\delta],\ \ \delta\ll 1.
\end{eqnarray}
We can find $\delta$ to leading order in $1/\lambda\tau$, by using the relation $\gamma'(-d)=\lambda_d$. The expansion of $\gamma(-\mu)$ near $\mu=d$
in the steady-state condition $\gamma(-\mu)=-\mu/\tau$ gives
\begin{eqnarray}&&
\gamma(-d+d\delta)\approx d \lambda_d\delta\approx -\frac{d}{\tau},\ \ \delta\approx -\frac{1}{\lambda_d\tau},  \ |\lambda_d|\tau\ll 1.\nonumber
\end{eqnarray}
Thus, well above the transition, the exponent $\mu$ depends  linearly on the inverse relaxation time (remember that $\lambda_d<0$),
\begin{eqnarray}&&
\mu=d+\frac{d}{\lambda_d\tau},\ \ |\lambda_d\tau|\gg 1.
\end{eqnarray}
Correspondingly, the steady state distribution of the polymer extension well above the transition satisfies
\begin{eqnarray}&&
P_{ss}(R)=\frac{1}{ N'' R_{max}}\left(\frac{R}{R_{max}}\right)^{d[1-|\lambda_d\tau|^{-1}]-1}.
\end{eqnarray}
For the Kraichnan model $|\lambda_d|=\lambda$, and the result from above holds generally.
This prediction gives the power-law distribution for polymers in a flow well above the coil-stretch transition. Both terms in the above expansion
constitute a new result that can be tested
experimentally, either by putting polymers with a large relaxation time in a given flow, or by considering a  vigorous flow with high $|\lambda_d|$
for a polymer with a given $\tau$.

\section{The distribution in the vicinity of the coil-stretch transition}

Another limiting case, opposite to the situation well-above the transition just described in the previous section, concerns the vicinity of the
coil-stretch transition,  for which $1/\tau\approx \lambda_1$. The corresponding result below the transition was found in \cite{BFL1}), and while
the generalization to the case above the transition is straightforward, we bring it here for completeness. It is clear from the stationarity condition (\ref{stcond}) that $\mu$ is close to zero
in this case. Expanding $\gamma(-\mu)$ near $\mu=0$ to second order, we find
\begin{eqnarray}&&
-\lambda_1 \mu+\frac{\gamma''(0)\mu^2}{2}=-\frac{\mu}{\tau}.
\end{eqnarray}
Discarding the trivial solution $\mu=0$, one has
\begin{eqnarray}&&
\mu=\kappa\left[1-\frac{1}{\lambda_1\tau}\right],\ \ \kappa\equiv\frac{2 \lambda_1}{\gamma''(0)}=\frac{2 \gamma'(0)}{\gamma''(0)}.
\end{eqnarray}
It is clear from $\gamma(-d)=0$ that $\kappa$, which is positive due to the convexity of $\gamma(\alpha)$, is estimated as $d$.
This agrees with the result of Ref. \cite{BFL1}, where it was used slightly below the coil-stretch transition. Here we only show that the same result
also applies slightly above this transition. We have
\begin{eqnarray}&&
\mu=\kappa-\frac{\kappa}{\lambda_1\tau},\ \ |1-\lambda_1\tau|\ll 1.
\end{eqnarray}
This clearly shows that the coil-stretch transition occurs at $\lambda_1\tau=1$, where $\mu$ changes
sign. The corresponding steady-state distribution of the polymer extension near the transition satisfies
\begin{eqnarray}&&
P_{ss}(R)=\frac{1}{ N'' R_{max}}\left(\frac{R}{R_{max}}\right)^{\kappa[1-(\lambda_1\tau)^{-1}]-1}.
\end{eqnarray}
For the Kraichnan model, we have $\gamma''(0)=2\lambda_1/d$, see Eq.~(\ref{gammaKraich}). Thus,  $\kappa=d$ and
$\mu=d[1-(\lambda\tau)^{-1}]$, in agreement with the previous results. The numerical confirmation of the result can be found in
\cite{MV,B}.

\section{Synchronization of polymers in the flow}

The synchronization of polymers in the flow is considered along the same lines as for threads. The modification of the motion equation for the difference
$\delta\rho$ of the radial variables in Sec. \ref{synchro} only concerns the addition of the term $-\delta \rho/\tau$,  which makes the relaxation of the
extensions to the same size even faster.

There is, however, one important new regime, that exists for polymers, but not for threads: the regime below the coil-stretch transition, where
$\lambda\tau<1$. There, the volume fraction of the regions, where polymers are stretched to a size not smaller than some $R$ satisfying
$R\gg R_{coil}$, is much smaller than unity and behaves as a positive power law in $R_{coil}/R$. Still, these regions are rather frequent in space,
 as their  decay is only a power law. The theory constructed here then shows that  the polymers are synchronized in these rare regions, where they are strongly stretched. It was shown in Ref. \cite{BFL1} that the rare fluctuations of the velocity gradient $\sigma$, that form $R$ much larger than the most probable size $R_{coil}$, persist during a long time-interval, where one can apply the dynamics
\begin{eqnarray}&&
\frac{d\bm R}{dt}=\sigma \bm R-\frac{\bm R}{\tau} .
\end{eqnarray}
This dynamics holds during a time-interval much larger than $(\lambda_1-\lambda_2)^{-1}$. Thus, these rare fluctuations of the flow, which stretch
the polymers, also synchronize these polymers by starting from initially uncorrelated coiled-up states.

\section{Synchronization and hydrodynamic equations of dilute polymer solutions}

The main application of the phenomenon of synchronization is to the hydrodynamic equations of dilute polymer solutions.
The synchronization implies that in the regions where the polymers are stretched, one can introduce a smooth macroscopic field $\bm R(\bm x, t)$, providing the end-to-end distances of polymers near the point $\bm x$. This field is defined by
\begin{eqnarray}&&
\bm R(\bm x, t)=\sum_i\bm R_i(t)\delta\left[\bm x-\bm x_i(t)\right],
\end{eqnarray}
where $i$ is the index of the polymer, and $\bm x_i(t)$ is the position of the molecule's center of mass. Differentiation yields
\begin{eqnarray}&&
\frac{\partial \bm R}{\partial t}+\bm u\cdot\nabla \bm R=\bm R\cdot\nabla u-\frac{\bm R}{\tau},
\end{eqnarray}
where one writes the substantial derivative of $\bm R(\bm x, t)$ instead of the ordinary time derivative appearing in the equation for $\bm R$ of a
single polymer. As it was observed in \cite{FL}, the vector field equation of the above form leads to the purely decaying equation $(\partial_t+\bm u\cdot\nabla) \nabla\cdot \bm R=-\nabla\cdot \bm R/\tau$ for $\nabla\cdot \bm R$. Thus, in the steady state one can always assume $\nabla\cdot \bm R=0$.
This conclusion on the spatial distribution of polymers seems to be non-trivial.

The above equations imply that, if the polymer density $n$ obeys the equation (we use $\nabla\cdot  \bm u=0$)
\begin{eqnarray}&&
\frac{\partial n}{\partial t}+\bm u\cdot\nabla n=\kappa\nabla^2 n,
\end{eqnarray}
where $\kappa$ describes the diffusion of polymers, then the field ${\tilde \Pi}_{ij}\equiv n R_iR_j$ obeys
\begin{eqnarray}&&\!\!\!\!\!\!\!\!
\frac{\partial {\tilde \Pi}}{\partial t}+\bm u\cdot\nabla {\tilde \Pi}={\tilde \Pi}\nabla \bm u+(\nabla \bm u)^t{\tilde \Pi}-\frac{2{\tilde \Pi}}{\tau}+\kappa R_iR_j \nabla^2 n.\nonumber
\end{eqnarray}
In many situations one can neglect the last term. In particular, this is true when polymers are uniformly distributed in space, which is a possible
steady-state solution. It should be mentioned that there are indications that in certain experimental situations the inhomogeneity of the polymers' distribution in space may play an important role. Postponing the consideration of such situations to future work, we assume that the last term is negligible. In that case, ${\tilde \Pi}_{ij}$ obeys a closed equation and, noticing that ${\tilde \Pi}$ is proportional to the polymer contribution $\Pi$ to the stress tensor (where the coefficient of proportionality includes the rigidity of a single molecule), we conclude that one can write a closed system of equations for $\Pi_{ij}$ and $\bm u$ (not including $n$). Furthermore, in the case $n=const$ one can directly write $\Pi_{ij}(\bm x, t)=B_i(\bm x, t)B_j(\bm x, t),$ where $\bm B(\bm x, t)$ is proportional to $\bm R(\bm x, t)$. Using $\nabla\cdot\bm R=0$, one finds the following hydrodynamic equations for dilute polymer solutions
\begin{eqnarray}&&\!\!\!\!\!\!\!\!
\partial_t\bm u+(\bm u\cdot \nabla)\bm u=-\nabla p+\nu \nabla^2\bm u+\bm B\cdot\nabla\bm B,\ \ \nabla\cdot\bm u=0,\nonumber\\&&\!\!\!\!\!\!\!\!
\partial_t\bm B+(\bm u\cdot \nabla)\bm B=(\bm B\cdot \nabla)\bm u-\frac{\bm B}{\tau},\ \
\nabla\cdot\bm B=0.
\end{eqnarray}
The above equations were introduced in \cite{FL}, based on the purely macroscopic consideration that the hydrodynamic equations admit the single-axis ansatz for $\Pi_{ij}$. The synchronization provides the microscopic basis, explaining the ansatz. It should be noted that, while the synchronization implies the single-axis ansatz, the converse is not true: one may have significant fluctuations in the end-to-end distances of individual polymers,
non-contradicting the single-axis ansatz for the molecule-averaged quantity $\Pi_{ij}$.

Thus,  synchronization clarifies the derivation of hydrodynamic equations for dilute polymer solutions, and can be used for a better  understanding
of the role of inhomogeneity in the polymer distribution in space.

\section{Conclusion}

In this work we derived new results on the behavior of small elastic objects in a turbulent flow. These results concern both the single particle
behavior, and the behavior of many particles, even though it is always assumed that there is no interaction between the particles. The results
can also be applied to chaotic flows satisfying the appropriate conditions of the decay of correlations.

\begin{figure}
\includegraphics[width=8.0 cm,clip=]{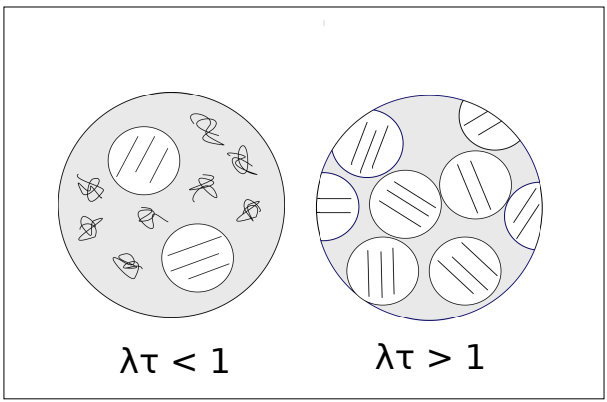}
\caption{Sketch of the domain structure below $(\lambda \tau < 1)$ and above
$(\lambda \tau > 1)$ the coil-stretch transition for polymers in a turbulent flow.
The characteristic size of the ordered domains, which spontaneously arise
below the coil-stretch transition, is the Kolmogorov length. Above the
transition, if polymer back reaction is important, an even larger correlation
length is possible \cite{BFL2,FL}.}
 \label{domains}
\end{figure}
We first consider the behavior of small threads. The threads do not resist the average stretching by the flow, unless they reach the size comparable
with their full length. Once this happens, the thread's size ''stabilizes'' at about the full length. The thread becomes similar to a rod, and it rotates in
the same way as the vector connecting two fluid particles in the same flow. This continues unless the thread arrives at a region of the flow,
where the local velocity gradient leads to a shrinking of the thread. The probability for the threads to shrink to a size
considerably smaller than the full length is significant: it is known to decay only as a power law of the thread's extension.
We show that the exponent of this power law is
universal and is given by $d-1$, where
$d=2, 3$ is the space dimension. Notably, the answer is the same for any statistics of the velocity field, for which the Lagrangian velocity gradient
has a finite correlation time.

One could think that this universality has a simple explanation: the answer would have the same form, if the second end of the thread would
perform a simple diffusion in a ball, centered at one end of the thread, with a radius given by the thread's length, and with a reflecting
inner surface. However, the physical motion is quite different: it is the logarithm of the distance that is driven by the noise and not the
distance itself. The noise has a non-zero average, in contrast to diffusion. Finally, the precise ratio between the average and the
dispersion of the noise, such as the one of the Kraichnan model, was shown to be important to get the correct answer.
This becomes clear when one passes to consider polymers, that resist stretching by the flow
 with harmonic forces for sizes much smaller than their full length. Here too, a power law is known to describe the probability of the polymer shrinking to
a size much smaller than the full length (it should be mentioned that this power law has a  cutoff at the characteristic equilibrium size $R_{coil}$ -
the coil's radius, where the thermal noise becomes important). However, this power law is known to be non-universal and to depend on
the velocity statistics (and the polymer's
relaxation time). Nonetheless, we show that the exponent of the power law has a more universal form in the limit well above the coil-stretch transition, where only one constant $\lambda_d$ completely characterizes the velocity statistics.

Our findings are in accord with the recent experimental findings in \cite{SL1} (see also \cite{SL2}), where
a universal exponent was obtained for the power-law of the PDF of the polymer size in a shear flow. The shear flow is different from the turbulent flow we study here, where small-scale isotropy
holds, and the generalization of our results to anisotropic situations such as shear, is postponed for future work.

Furthermore, we show that different particles (polymers or threads) get synchronized by the flow that stretches them.
Our conclusion concerning the synchronization of polymers has particular importance in view of the physical relevance for polymer solution.
Below the coil-stretch transition, the polymers are coiled in most of the space, and the properties of the solution are very close to those
of the solvent. The configurations of different coiled-up polymers get correlated because the same flow tries to stretch them, but the correlation is
negligibly small. However, there are also regions where polymers
are stretched to $R\geq R_0$, where $R_0$ is some number much greater than $R_{coil}$. The probability of such regions decays only as a power of
$R_{coil}/R_0$, and  so these regions occur in space rather frequently. The end-to-end distances
in such regions are  created by velocity-gradient fluctuations, for which the thermal noise is  negligible during time-scales much larger than those needed for
synchronization. Thus, in these regions the end-to-end distances of different polymer molecules are nearly
identical (of course, the size of these regions is much smaller than the Kolmogorov length). As the Lyapunov exponent of the flow grows,
the "stretched regions" become more frequent in space, and above the transition they fill almost all the space. See the sketch of such a situation in
Fig.  \ref{domains}. Thus, above the transition, there is a
well-defined macroscopic field $\bm R(t, \bm r)$ that varies over the scale of smoothness of the flow, and to which we also refer as domains.
This field was originally introduced in Ref. \cite{FL} as a
consequence of the hydrodynamic equations. Here, we give a microscopic meaning and interpretation to that field and discuss the implications
for the hydrodynamic equations.
The above predictions can be tested experimentally.

We assumed in our analysis that the flow velocity is independent of the configurations of the polymers, so the stretching of the polymers is arrested by nonlinear elasticity. However, the most interesting case occurs when the stretching is arrested by the polymers' back reaction on the flow. Then
there is a size $R_{back}<R_{max}$ such that, when polymers are extended to that size, their back reaction on the flow becomes significant.
Formally, the back reaction is realized via the polymer contribution to the momentum
stress tensor. For example, if the polymer concentration is sufficiently large, then $R_{back}\ll R_{max}$, and one can apply the approximation of harmonic forces, where the contribution is proportional to $R_iR_j$, see \cite{BFL2,FL}.
One can see that our derivation of the power law still applies at $R\ll \min[R_{back}, R_{max}]$, where $\bm R$ decouples from $\sigma$ by the
definition of $R_{back}$, and our considerations can be repeated. As to synchronization, it is likely to hold without changes, but detailed study is needed.
The first question would be, whether the orientations of different polymers still satisfy the same equation as the orientations of the end-to-end vectors
of two fluid particles. If so, then -- assuming there are no anomalies and the Jacobi matrix behaves in the usual way characteristic for chaos that was
described above -- the orientations will get equal, and one expects synchronization to hold. This conclusion is immediate if one could model the
effect of back reaction on polymers by an effective radial resistive force that becomes important at $R\sim R_{back}$. The detailed study of whether our conclusions can be transferred to the important case of flows with back reaction, that includes the drag reduction problem, is the subject of
future work.


%

\section{Acknowledgments}

Our work was supported by the Society of Austrian Friends of the University of Tel Aviv.


\begin{references}

\bibitem{BFL1} E.~Balkovsky, A.~Fouxon, and V.~Lebedev, Phys. Rev. Lett.
{\bf 84}, 4765 (2000).

\bibitem{Chertkov} M. Chertkov, Phys. Rev. Lett. {\bf 84}, 4761 (2000).

\bibitem{Th} J.-L. Thiffeault, Phys. Lett. A {\bf 308}, 445–450 (2003).

\bibitem{Celani} A. Celani, S. Musacchio, and D. Vincenzi, J. Stat. Phys., {\bf 118}, Nos. 3/4, (2005).

\bibitem{Afonso}  M. Afonso and D. Vincenzi, J. Fluid Mech. {\bf 540}, 99 (2005).

\bibitem{St} S. Gerashchenko, C. Chevallard, and V. Steinberg, Europhys. Lett., {\bf 71}, 221–227 (2005).

\bibitem{GW} T. Watanabe and T. Gotoh, Phys. Rev. E {\bf 81}, 066301 (2010).

\bibitem{review} G. Falkovich, K. Gawedzki, and M. Vergassola, Rev. Mod. Phys. {\bf 73}, 913–975 (2001).

\bibitem{SL1} Y. Liu and V. Steinberg, EPL, Europhys. Lett., {\bf 90}, 44005 (2010).

\bibitem{Lumley1} J. L. Lumley, J. Polym. Sci. Macromol. Rev. {\bf 7}, 263 (1973).

\bibitem{Lumley2}  J. L. Lumley, Symp. Math. {\bf 9}, 315 (1972).

\bibitem{deGennes} P. G. de Gennes, J. Chem. Phys. {\bf 60}, 5030 (1974).

\bibitem{BFL2} E. Balkovsky, A. Fouxon, and V. Lebedev, Phys. Rev. E {\bf 64}, 056301 (2001).

\bibitem{FL} A. Fouxon and V. Lebedev, Phys. Fluids {\bf 15}, 2060 (2003).

\bibitem{Batchelor} G. K. Batchelor, J. Fl. Mech. , {\bf 46}, 813 (1971).

\bibitem{GB} A. Gyr and H.-W. Bewersdorf, {\it Drag Reduction in Turbulent Flows by Additives} (Kluwer, London, 1995).

\bibitem{JC} S. Jin and L. R Collins, New J. Phys. {\textbf 9},  360 (2007).

\bibitem{BF} E.~Balkovsky and A.~Fouxon, Phys. Rev. E {\bf 60}, 4164 (1999).

\bibitem{Risken} H. Risken,  {\it The Fokker-Planck Equation: Methods of Solutions and Applications} (Springer, Berlin, 1996).

\bibitem{Kivotides} D. Kivotides, S. L. Wilkin, T. G. Theofanous, Phys. Lett. A {\bf 375}, 48–52 (2010).

\bibitem{Oseledets} V. I. Oseledets, Trans. Mosc. Math. Soc. {\bf 19}, 197 (1968).

\bibitem{Zeldovich} Ya. B.  Zeldovich, A. A.  Ruzmaikin,
S. A.  Molchanov and D. D. Sokoloff, J. Fluid Mech. {\bf 144}, 1-11 (1984).


\bibitem{Bof} G. Boffetta, A. Celani, and S. Musacchio, Phys. Rev. Lett. {\bf 91}, 034501 (2003).

\bibitem{MV} S. Musacchio and D. Vincenzi, J. Fluid Mech. {\bf 670}, 326 (2011).

\bibitem{B} F. Bagheri, D. Mitra, P. Perlekar, and Luca Brandt, arXiv:1011.3766.

\bibitem{SL2} Y. Liu and V. Steinberg, EPL, Europhys. Lett., {\bf 90}, 44002 (2010).

\end{references}
\end{document}